\documentclass[12pt,preprint]{aastex}

\shorttitle{Galaxies in 2PIGG groups}
\shortauthors{Robotham et al.}

\begin{document}


\title{Galaxy luminosities in 2PIGG Groups}


\author{Aaron Robotham, Christopher Wallace, Steven Phillipps, 
Roberto De Propris\altaffilmark{1}}
\affil{H.~H.~Wills Physics Laboratory, University of Bristol,\\
       Tyndall Avenue, Bristol, BS8 1TL, United Kingdom}
\email{A.Robotham@bristol.ac.uk}

\altaffiltext{1}{Cerro Tololo Inter-American Observatory, Casilla 603, La Serena,
Chile}




\begin{abstract}

We derive composite luminosity functions (LF) for galaxies in groups and examine 
the behaviour of the LF as a function of group luminosity (used as an indicator of
group or halo mass). We consider both the entire galaxy population and split 
galaxies into red and blue (quiescent and star forming) samples, in order to 
examine possible mechanisms behind observed variations of galaxy properties with 
environment. We find evidence that $M^*$ brightens and $\alpha$ steepens with 
group luminosity, until a threshold value where the LF parameters stabilize at 
those found in rich clusters. The effect is seen in the total LF and for the 
blue and red galaxies separately. The behaviour of the quiescent and star-forming 
samples is qualitatively consistent with variations resulting from
interactions and mergers, where mergers build the bright end of the luminosity
function at the same time as dwarf irregulars have their star formation quenched
and evolve into dwarf ellipticals. These processes appear to take place
preferentially in low luminosity groups and to be complete at a group luminosity
of $-22.5$ in $B$, corresponding to a halo mass of order $10^{13.5}\ M_{\odot}$.

\end{abstract}




\section{Introduction}

There is considerable evidence that most galaxy properties, such as morphology 
\citep{dressler80,postman84,allington93}, star formation rates \citep{lewis02,gomez03,
balogh04}, color \citep{depropris04, kauffmann04} and even luminosity function 
parameters (Driver, Couch \& Phillipps 1998; \citealt{phillipps98, croton05}) 
correlate well with local density. It is likely that the main mechanism behind 
the observed correlations resides in rapid cessation of star formation in galaxies, 
once a threshold local density is reached (e.g. \citealt{balogh04}). Observations 
of the color-magnitude relation in the neighborhood of clusters also point to 
a swift change in galaxy colors once the density in the surrounding structure 
reaches a critical value \citep{kodama01,kodama05,tanaka05}. The environment where 
the transition in galaxy properties occurs, from those of typical field regions 
to those seen in dense clusters, is an important indicator of the evolutionary 
processes involved (see e.g. \citealt{pimbblet06}, and references therein).

It is generally believed that most galaxy transformations (in morphology, star
formation rates and color) take place in groups, since these are the environments
where most galaxies reside and groups have sufficiently high densities, but 
sufficiently low velocity dispersions, that interactions and mergers may take 
place (e.g. \citealt{ghigna98,mulchaey98}). Indeed, compact groups are seen to
populate the infall regions of the rich clusters studied by \cite{lewis02} and 
\cite{tanaka05}, where the suppression of star formation and galaxy evolution 
appear to be taking place. Other processes, such as ram pressure stripping 
are expected to be important only at higher densities (e.g. Abadi, Moore \&
Bower 1999).

\cite{kauffmann04}, \cite{depropris04} and \cite{croton05} show that most of 
the galaxies forming stars at the present epoch have relatively low 
luminosities and consist mostly of dwarf spirals and irregulars, while 
giants are mostly quiescent and tend to reside in the denser environments 
\citep{yang05a}. This suggests that a study of the behavior of dwarf (and 
intermediate luminosity) galaxies in the group environment, together with 
a comparison of star-forming and quiescent galaxies may provide useful 
insight into the mechanisms that transform field galaxies into typical 
cluster members.

One of the most versatile (if simplistic) tools for studies of galaxy 
population variations  is the galaxy luminosity function (LF). The 
characteristic magnitude $M^*$ provides a measure of the variation of 
giant galaxy luminosity or mass, while the faint end slope $\alpha$ 
can be related to the properties of the dwarf population. LFs in different 
wavebands, or selected according to appropriate color cuts, may be used 
to understand the stellar populations and star formation histories 
of the galaxy population under study.

In this paper we consider the variation of the galaxy LF in a sample of nearby 
groups spanning a wide range of masses.  This topic was previously investigated 
by \cite{christlein00}, using Las Campanas Redshift Survey data (though this survey 
is now known to suffer from significant surface brightness selection effects), and 
by \cite{eke04a}, who looked at three, dynamically determined, mass bins. Here, 
we consider the entire population, split this into several group luminosity bins
and consider quiescent (red) and star-forming (blue) galaxies separately, in order
to explore the possible roles of mergers and star formation suppression. We mainly 
concentrate on the behavior of lower luminosity galaxies, i.e. on the faint end of the LF, 
as a function of group properties, as this is where we expect to see the strongest signature of
evolutionary effects. Dwarfs are believed to be fragile systems, whose properties are easily 
affected by dense environments (e.g. \citealt{moore96}). 

We start by describing the selection of groups and the derivation of the LF parameters 
in Section 2, and in Section 3 we discuss the observed trends and present possible 
interpretations  of our findings. We adopt a cosmology with $\Omega_M=0.3$ and 
$\Omega_{ \Lambda}= 0.7$ (though this has minimal impact on our results) and normalize 
distances to H$_0=100$ km s$^{-1}$ Mpc$^{-1}$ (i.e. all quoted magnitudes should be 
read as $M - 5 \log h$, where $h = H_0/100$ km s$^{-1}$ Mpc$^{-1}$).

\section{Methodology}

The largest publicly available group sample is the 2dF  Percolation-Inferred Galaxy 
Groups (2PIGG) catalog \citep{eke04a}, selected from the 2dF Galaxy Redshift
Survey (2dFGRS -- \citealt{colless01}). We begin by selecting all  groups from this 
catalogue with $0.05 < z < 0.10$. The $z=0.05$ lower limit is motivated by the small 
volume sampled by 2PIGG at low redshift (so that groups may not be representative), 
while at $z=0.10$ the 2dFGRS apparent magnitude limit begins to exclude even moderate 
luminosity galaxies ($M_B < -18$) from the sample.

Since 2PIGG includes groupings as small as 2 galaxies, we also impose a minimum size
of 5 members to the groups we study. This should ensure that the groups we select are 
likely to be actual bound physical systems. These choices leave a total of 1535 
groups containing a total of 21,752 galaxies (when corrected for completeness, as 
described below).

Clearly the fainter limiting absolute magnitude for the nearer groups means that
less rich groups (in terms of giant members) can be included at lower $z$.
For instance, we could in principle include groups entirely composed of dwarfs.
However to minimise any effects due to this, we use only galaxies bright enough 
to be visible out to the edge of our survey volume ($M_{b_j} < -18.6$) in the 
estimation of the group masses. On the other hand, exploiting the nearer groups 
allows us to trace galaxies $\sim 1$ magnitude further down the LF into the dwarf 
regime and hence sample some poorer groups. In fact, using only galaxies brighter than
$M_{b_j}=-18.6$ throughout makes no noticeable difference to our results,
except in worsening the statistics for the smaller groups.

We calculate absolute $M_{b_J}$ magnitudes and rest-frame $b_J-r_F$ colors 
for galaxies in these groups following the prescriptions of \cite{cole05}: 
we ignore the relatively small amount of evolution within the narrow redshift
range we survey, as this correction is both small and somewhat uncertain. We 
calculate the group luminosity, expressed as a magnitude $M^G_{b_J}$,
by summing up galaxies brighter than $M_{b_J}=-18.6$ (the faintest galaxy we can 
detect at the highest redshift we consider), with weights determined by the 
survey spectroscopic completeness, as detailed in \cite{eke04b}. This $B$ band group 
luminosity is then used as our primary indicator of group baryonic mass and thus as a 
proxy for the group halo mass (see e.g.\citealt{yang05b}). Note that our measure is 
very close to that of \cite{yang05a}, who use all galaxies brighter than $M_B = -18$ to 
determine group luminosities. We discuss the question of the most suitable environmental
measures and proxies further in section 3. We note here, though, that our lowest
mass group environment will not be synonymous with the generic `field' environment.
The latter is an average of groups over a whole range of (low) multiplicities
and masses.

We split our sample of groups into 10 bins of increasing total luminosity within 
the range equivalent to $M^G_{b_J}=-20$ to $M^G_{b_J}=-24$.  Bin widths are chosen 
so that each bin contains approximately the same total number of galaxies. This 
provides a much finer subdivision than used by \cite{eke04b} and allows us the opportunity 
to search in more detail for any trends and/or transition in properties from cluster-like 
to field-like environments. For each of the luminosity bins we compute a composite galaxy LF, 
following the method described by \cite{colless89}. In addition to the LFs derived 
for the entire sample we also create LFs for red (quiescent) and blue (star-forming) galaxies 
separately, adopting a color of $b_j-r_F=1.07$ at which to divide the two samples 
\citep{cole05}.

Table 1 shows the derived values of the LF parameters, together with marginal 1$\sigma$
errors. Most of the changes in $M^*$ and $\alpha$ appear to occur in the six lowest
luminosity bins, which we will focus on in our discussion. We plot the LFs for these
six bins (for all, blue and red samples) in Figures 1 (all), 2 (blue galaxies) and 3
(red galaxies) and the related error ellipses in Figure 4.     

\section{Results and Discussion}

Figure 5 summarizes the variation in LF parameters with group luminosity for each 
of the samples we consider. As in \cite{eke04b} we find two noticeable trends in 
the data as a whole, viz. that $M^*$ brightens and $\alpha$ steepens with increasing 
group mass. With the finer resolution we adopt, we are able to see that, in fact, 
the LF parameter values appear to reach approximately constant levels, comparable to 
those seen in  rich clusters \citep{depropris03}, for groups brighter than $M^G_{b_J} 
\simeq -22.5$. This is similar to the behaviour reported by \cite{dominguez02} and
\cite{martinez02} from their analysis of groups in the 2dFGRS 100K galaxy release 
(with a different grouping algorithm) and is presumably related to the observation 
that galaxy properties appear to change at some `threshold' density in the 
neighborhood of clusters \citep{kodama01,lewis02,tanaka05}. Of course, once we 
reach these asymptotic values, the group luminosity which we use as a discriminator 
will directly mirror the richness of the group.

Consider each of these trends in turn. As noted above, the characteristic magnitude 
$M^*$ levels off (or may even go fainter again) for groups with magnitudes brighter 
than $M^G_{b_J} \simeq -22$ ($\simeq 10^{11} h^{-2} L_{\odot}$). The value seen for 
these more luminous groups is similar to that seen in rich clusters, even though they 
start from only moderate mass groups; assuming $h=0.7$ and a corresponding average 
universal ${\cal M}/L$ ratio of 300 in solar units (e.g. \citealt{bahcall00}), they 
have group (halo) masses ${\cal M}_h \simeq 6 \times 10^{13}$ to $3 \times 10^{14} 
{\cal M}_{\odot}$. We obtain a similar value for the mass at the `turnover' point,
${\cal M}_h \simeq 10^{13.6}$, if we use Eke et al.'s (2004b) luminosity
dependent mass-to-light ratios. 

Among the less luminous groups, there is a clear faintening of $M^*$ as one goes
from intermediate to low luminosity. The dimming is seen in each left hand panel 
of Figure 3, though it is less clear for the blue galaxies. Assuming a simple 
linear relation we find slopes in $M^*$ vs. $M^G_{b_J}$, calculated as bisector
fits of $0.77 \pm 0.09$ for the whole sample, $0.63 \pm 0.12$ for the red galaxies
and $0.28 \pm 0.13$ for the blue galaxies, with goodness of fit coefficients $R^2$ 
of 0.983, 0.960 and 0.717, respectively.

One could consider that this is merely a selection effect due to the enforced 
absence of very luminous galaxies: clearly we can not have an $M_{b_J} = -21$ galaxy 
in an $M^G_{b_J} = -20$ group. However, we would argue that it is still an interesting 
and informative exercise to determine how the available baryonic mass in a low mass
group {\it does} organise itself (i.e. the `conditional' LF; see e.g. van den Bosch,
Yang \& Mo 2003). From Figure 1, it is clear that the mass in the small groups is 
still shared out in a Schechter function-like distribution, but with a steadily declining 
bright end characteristic luminosity as we go to smaller groups\footnote{\cite{degrijs06}
present an interesting discussion of the related question of how the mass function
of the stars in small star clusters is `filled-up' until the available mass is used}.

We have also tested the `null' hypothesis of selecting out groups with low total
luminosities from sets of galaxies (mock groups) produced by statistically sampling
a fixed LF (both the 2dFGRS cluster LF of \citealt{depropris03} and the field LFs
of \citealt{croton05}). As expected, $M^*$ becomes fainter for the mock groups 
with low total luminosity (due to the obvious lack of bright galaxies). However,
these mock groups then turn out to be overpopulated with fainter objects compared
to the real groups, leading to a steepening, not flattening, of the faint end 
slope as we go to lower mass groups. This gives us further confidence that the
changes in slope we observe are real physical effects. It is more difficult to
disentagle whether the change in $M^*$ is due to the effect of sampling the LF
or is a real physical effect. However, though for a given small multiplicity 
group it is unlikely that a bright galaxy will be selected from a given purely 
statistical representation of the LF, once we sum over many groups, as here, the 
original LF will be restored (i.e. by chance, a few groups {\it will} have a bright member).

Note that \cite{hansen05} also find a shallower slope and fainter $M^*$ for their 
SDSS `clusters' with a low number of bright, red members. These objects are selected differently
from our groups (they do not use spectroscopic redshifts, as we do), but their Figure 10 
shows the same `saturation' of LF parameters for more massive systems as we find in our study.

Indeed, the observed behaviour is in excellent agreement with that predicted by the 
halo occupation model of \cite{yang05d}. At low group masses (which, as here, they 
use group luminosity to represent), they find that the luminosity of the brightest 
galaxies should increase as ${\cal M}_h^{2/3}$, with a much slower increase 
(slope $\simeq 1/4$) for more massive groups. Moreover, they predict that the 
turnover in slope should occur at group halo masses a few times $10^{13} M_{\odot}$, 
again in good agreement with our interpretation above (see also \citealt{vale04,cooray04}). 
\cite{yang05d} argue that this change in behaviour occurs at the transition from
efficient to inefficient cooling in different mass halos, and is the same
feature required in semi-analytical modelling of galaxy formation in order 
to match the bright end of the overall galaxy LF (see e.g. \citealt{benson03}). 
\cite{eke05} find that the group mass-to-light ratios (in the $K$ band) as 
a function of group mass level off at this same point, while in the semi-analytic
models of \cite{delucia06} there is a trend for early type galaxies in 
halos above this mass to be older, redder and metal richer (as against
bluer, younger and metal poorer in the lower mass halos).

As noted earlier, the changes appear more clearly and are larger in amplitude for the 
red galaxies: perhaps unsurprisingly, small groups are even more deficient in bright
red galaxies than bright blue ones. Even so, the generally similar trends for
quiescent and star forming galaxies imply that the brightening of $M^*$ in larger groups
is primarily a result of increasing galaxy masses, rather than just a variation in star 
formation and therefore in ${\cal M}/L$. A reasonable interpretation would then be that 
what we are seeing is a direct result of hierarchical growth of halos and the merging 
of galaxies within them (e.g. \citealt{dubinski98}; see also \citealt{vandokkum05,bell06}). 

Now consider the LF slope parameter $\alpha$. Unlike $M^*$, this is more descriptive 
of the moderate to low luminosity (dwarf) galaxy population (though we should note 
that we do not sample very deep into the dwarf regime, so will be insensitive to any 
turn-up, as seen in some rich clusters (e.g. \citealt{driver94,depropris95}; Smith, 
Driver \& Phillipps 1997) and X-ray bright groups \citep{miles04}, below $M_B \simeq -17$).
The slope should then be less influenced by the enforced absence of giant galaxies 
in the smaller groups. Nevertheless we again see a trend with group luminosity, with 
steeper slopes for the larger groups (as in \citealt{zabludoff00}), and again a levelling
off to the values observed for clusters once we reach $M^G_{b_J} \simeq -22$. 
Notice that even though $\alpha$ and $M^*$ will be linked by the fitting process, 
we generally sample to at least 2 magnitudes below $M^*$, so $\alpha$ should genuinely 
measure the faint end slope. Nevertheless, this parameter linkage may account for the 
unusually small value of $\alpha$ in the lowest group luminosity bin, where we have 
the smallest available fitting range. 

Recall that the low luminosity groups are not equivalent to the general field but
are a more specific environment. It is therefore not necessary (and indeed it is
not the case) that the faint end of the LF in such groups should match, or asymptotically
tend to, that of the field.  Field, and possibly void, LFs tend to have $\alpha \sim -1.2$
in the 2dF and SDSS studies, rather than our shallower slope. However, the two environments
are not comparable, because local densities are computed over much larger volumes than
the typical field size. Attempting to accommodate all the LFs with a fixed
$\alpha=-1.1$ results in poor fits for the low luminosity groups, but even then we
still see systematic changes in $M^*$ with group luminosity.

Again, this behaviour matches well with that prediced by \cite{yang05d} from 
the halo occupancy statistics (see also \citealt{cooray05}). They find flatter 
slopes ($\alpha \geq -1$) for halos below about $10^{13.5} M_{\odot}$ and steeper
ones above that, consistent with both the direction and the crossover point we 
see in our variation of $\alpha$ with group luminosity. 

The blue galaxies show a variation of $\alpha$ from about $-1.1$ to $-1.5$ (or even steeper),
as previously observed for many clusters compared to the field (e.g. \citealt{smith97}).
The red galaxies show an even greater variation, the lowest luminosity groups showing 
very clearly declining LFs ($\alpha > -1$) at the faint end. This argues for a general 
shortage of dwarf early type galaxies in these low density environments,
consistent with their status as the most highly clustered of all galaxy types 
\citep{ferguson91}.

Now consider the physical interpretation of the variation in faint end slope. If we are
assuming that the larger groups form by the hierarchical build up of smaller ones,
then we might look to the same sort of merger processes that we invoked earlier
in respect to the brightening of $M^*$ (c.f. \citealt{gonzalez05}). As we would 
expect mergers to take place most easily between the largest galaxies (e.g. \citealt{
makino97}), then the bright end of the LF should build up at the expense of intermediate 
luminosities (in any case, only such major mergers would substantially change the 
brightness of a large galaxy). This will then deplete the mid-section of the LF (c.f. 
\citealt{miles04}) and therefore create a steeper faint end, by default. However, this can not
be the whole picture, as the lowest luminosity groups have, in particular,
rather few faint red galaxies to start with; the change in slope of the red
galaxy LF is substantially greater than that of the total sample. We therefore
need to create extra faint red galaxies, the obvious source being intermediate
to low luminosity late type galaxies fading and reddening due to the cutting-off
of their star formation \citep{lin83,davies88,christlein03}. This then reinforces the 
idea of reduced star formation rates in increasingly dense environments \citep{lewis02,
depropris03,gomez03,depropris04,balogh04}.

The steepening of the LF for blue galaxies at around $M^G_{b_J} = -22.5$ is perhaps
surprisingly abrupt, but it is reminiscent of earlier claims for very steep LFs in 
clusters of galaxies \citep{driver94,depropris95}, especially in their outskirts 
\citep{phillipps98,boyce01}. Conceivably this effect may be due to star formation continuing 
in fainter dwarfs when it has ceased in intermediate mass ones (mimicking the global 
`downsizing' picture of cosmic star formation; Cowie et al. 1996). Another intriguing possibility would be that the blue galaxies represent a new population of tidal dwarfs (\citealt{mirabel92}; 
Hunsberger, Charlton and Zaritsky 1996), which tend to have steep luminosity functions
(Hunsberger, Charlton \& Zaritsky 1998), though they are not usually thought to be
as bright as the objects in our LFs. For the largest groups there may be a modest flattening
of the slope again, perhaps reflecting the smaller number of dwarfs seen in some
very dense environments \citep{boyce01}.

When attempting to interpret the observed trends, we should also consider the
appropriateness of our measure of the group or cluster environment. A group's
total mass is evidently a critical measure, as halo mass is the key
variable in structure formation models (\citealt{press74} {\it et seq.}; 
\citealt{yang05b}). The total luminosity has been demonstrated to be a good measure of 
total stellar mass and total halo mass \citep{padilla04,yang05c}. One could argue that 
the total $R$ band luminosity would be preferable to the $B$ band luminosity, 
as it is less subject to contributions from short term star formation, and that the 
$K$ band would be better still; $K$ band LFs should also more closely reflect the 
mass distribution of the individual group galaxies (e.g. Cole et al. 2001). 
In fact, $B$ and $R$ group luminosities turn out to be well correlated so
experiments dividing the sample by group $R$ magnitudes presents the same picture
as that already discussed, while $K$ band data is not yet available for a large
enough sample to overcome statistical uncertainties (though see \citealt{eke05}
for a current analysis).

We could also divide the groups purely by multiplicity, though this
will bring in additional (or at least different) selection effects, and theoretically
is expected to have wide, poissonian, variations for a given halo mass (e.g.
\citealt{kravtsov04,yang05d}), or by a dynamical measure such as velocity dispersion 
\citep{christlein00,eke04b} or parameters such as X-ray flux \citep{miles04}. We defer
a discussion of this and of a wider range of group properties to a subsequent paper.

To sum up, the results of our study suggest that changes in the galaxy population take 
place mostly among the lower luminosity groups and are consistent with the changes being 
brought about by merger/interaction-induced accretion and star formation suppression. 
We have argued that the observed changes in LF parameters are not merely a selection
effect due to the total available luminosity. 

The processes appear to be largely complete once the group luminosity reaches a threshold 
value $M^G_{b_J} \simeq -22.5$ (with projected densities of a few giant galaxies 
per Mpc$^2$).  This is similar to earlier observations in groups and 
clusters \citep{kodama01,dominguez02,martinez02,kodama05,sabatini05,tanaka05}. 
The `threshold' luminosity may arise during the growth of groups, when the internal velocity 
dispersion becomes so large as to inhibit further interactions despite the increased 
galaxy density. If the more massive groups then coalesce to produce clusters (as in
the hierarchical picture), the remarkable homogeneity in the properties of cluster 
galaxies observed by \cite{christlein03} and \cite{depropris03} would result naturally.

\section*{Acknowledgments}

We would like to thank the anonymous referee for his informative report.
AR and RDP acknowledge funding through the UK Particle Physics and Astrophysics
Research Council (PPARC).




\clearpage

\begin{deluxetable}{ccccccc}
\tablecaption{Luminosity Function parameters for groups}
\tabletypesize{\tiny}
\tablewidth{\textwidth}
\tablehead{
\colhead{$M^G_{b_J}$} &
\colhead{All Galaxies $M^*$} &
\colhead{All Galaxies $\alpha$} &
\colhead{Blue Galaxies $M^*$} &
\colhead{Blue Galaxies $\alpha$} &
\colhead{Red Galaxies $M^*$} &
\colhead{Red Galaxies $\alpha$}}
\startdata
$-20.02 \pm 0.47$ & $-18.71 \pm 0.10$ & $-0.62 \pm 0.10$ & $-19.11 \pm 0.20$ & $-1.00 \pm 0.17$ & $-18.63 \pm 0.12$ & $0.00 \pm 0.15$ \\
$-20.82 \pm 0.12$ & $-19.32 \pm 0.13$ & $-0.88 \pm 0.09$ & $-19.35 \pm 0.17$ & $-1.10 \pm 0.14$ & $-18.92 \pm 0.12$ & $-0.01 \pm 0.12$ \\
$-21.18 \pm 0.10$ & $-19.54 \pm 0.12$ & $-0.87 \pm 0.09$ & $-19.68 \pm 0.22$ & $-0.96 \pm 0.13$ & $-19.31 \pm 0.11$ & $-0.38 \pm 0.11$ \\
$-21.51 \pm 0.09$ & $-20.01 \pm 0.16$ & $-1.15 \pm 0.10$ & $-19.71 \pm 0.20$ & $-1.14 \pm 0.12$ & $-19.72 \pm 0.15$ & $-0.78 \pm 0.12$ \\
$-21.80 \pm 0.09$ & $-20.08 \pm 0.19$ & $-1.03 \pm 0.10$ & $-19.49 \pm 0.18$ & $-0.68 \pm 0.14$ & $-19.89 \pm 0.17$ & $-0.83 \pm 0.11$ \\
$-22.12 \pm 0.09$ & $-20.28 \pm 0.24$ & $-1.21 \pm 0.10$ & $-19.74 \pm 0.22$ & $-1.19 \pm 0.12$ & $-20.14 \pm 0.25$ & $-0.92 \pm 0.12$ \\
$-22.47 \pm 0.12$ & $-19.92 \pm 0.21$ & $-1.03 \pm 0.13$ & $-20.45 \pm 0.45$ & $-1.80 \pm 0.12$ & $-19.75 \pm 0.18$ & $-0.73 \pm 0.15$ \\
$-22.95 \pm 0.16$ & $-20.36 \pm 0.34$ & $-1.35 \pm 0.16$ & $-20.00 \pm 0.32$ & $-1.83 \pm 0.14$ & $-20.26 \pm 0.33$ & $-1.16 \pm 0.15$ \\
$-23.53 \pm 0.19$ & $-19.53 \pm 0.30$ & $-0.92 \pm 0.20$ & $-19.87 \pm 0.38$ & $-1.57 \pm 0.18$ & $-19.32 \pm 0.29$ & $-0.63 \pm 0.21$ \\
$-24.33 \pm 0.32$ & $-19.66 \pm 0.48$ & $-1.07 \pm 0.23$ & $-19.82 \pm 0.59$ & $-1.46 \pm 0.30$ & $-19.44 \pm 0.42$ & $-0.81 \pm 0.30$ \\
\enddata
\end{deluxetable}

\clearpage



\begin{figure}
\epsscale{0.95}
\plottwo{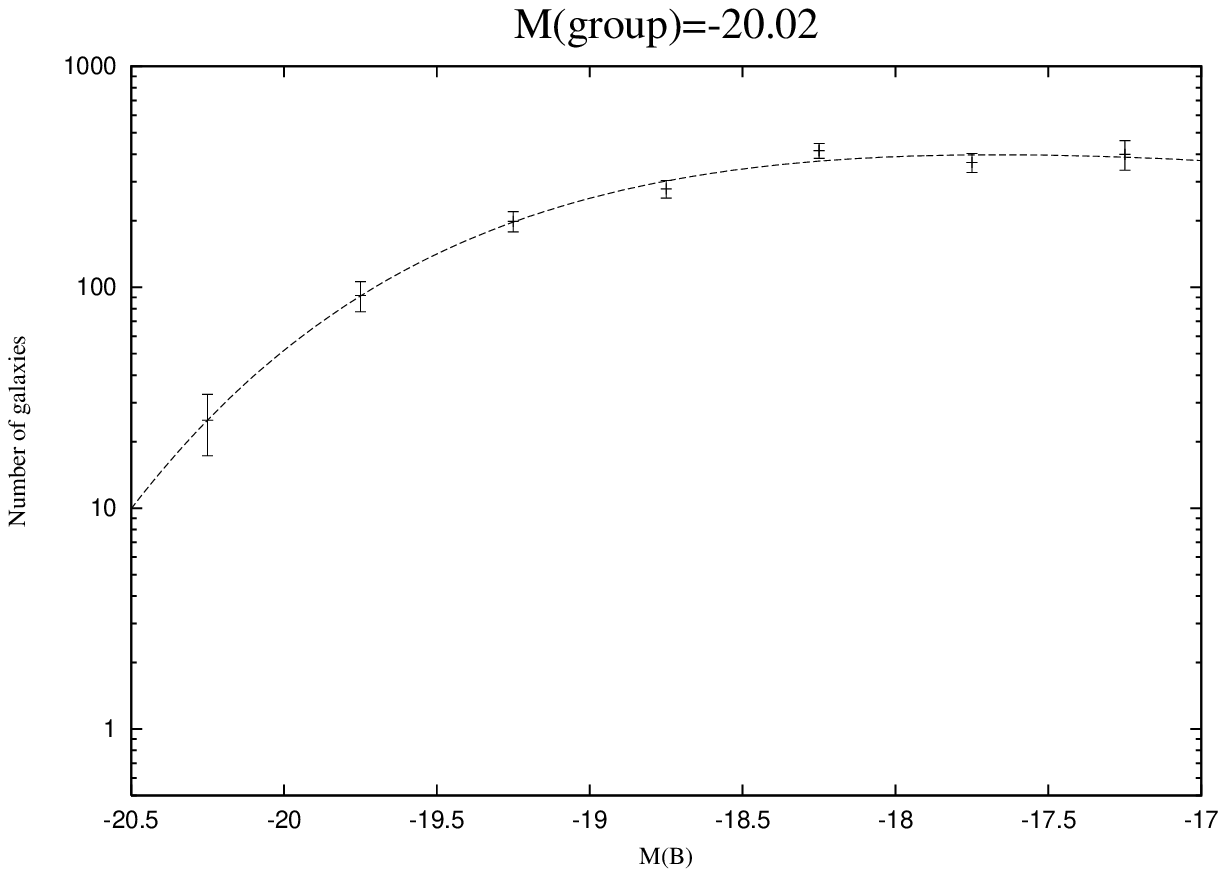}{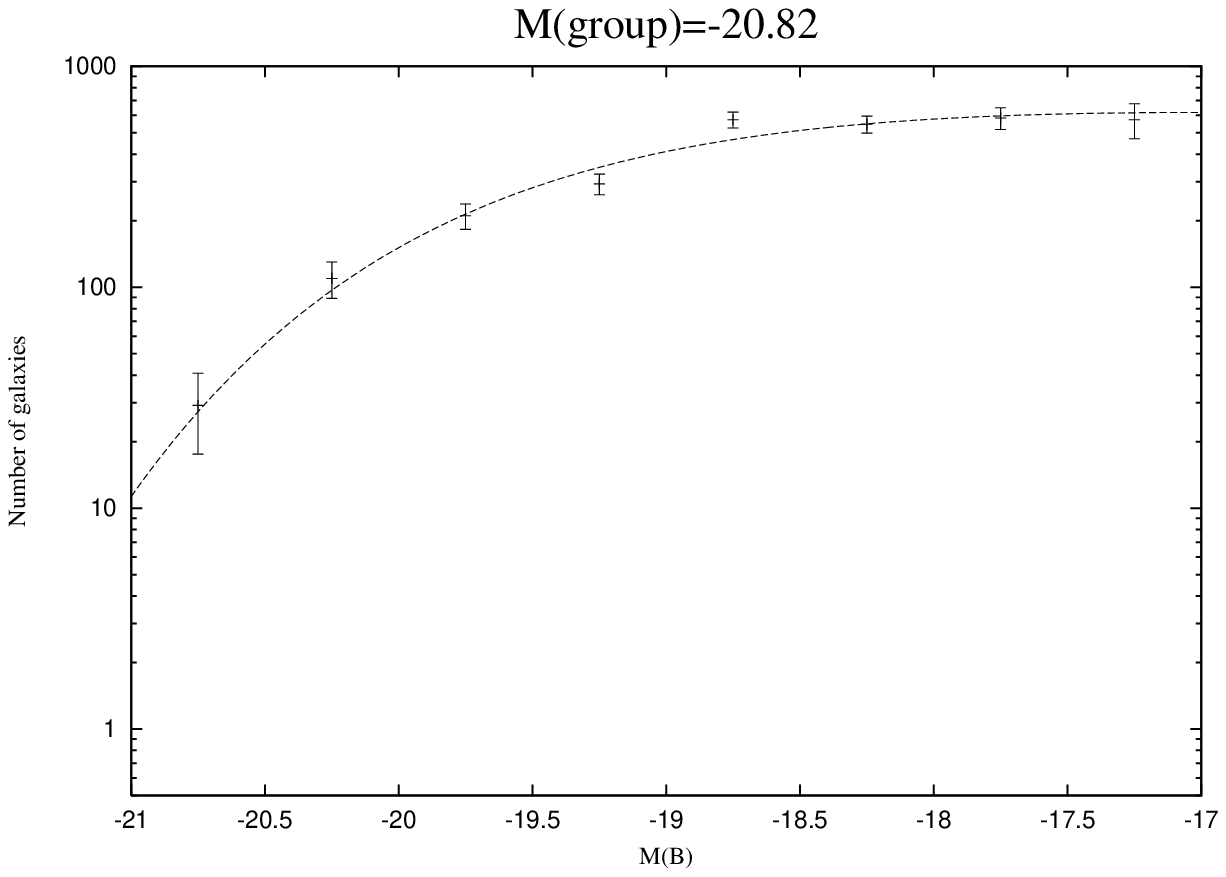} \\
\plottwo{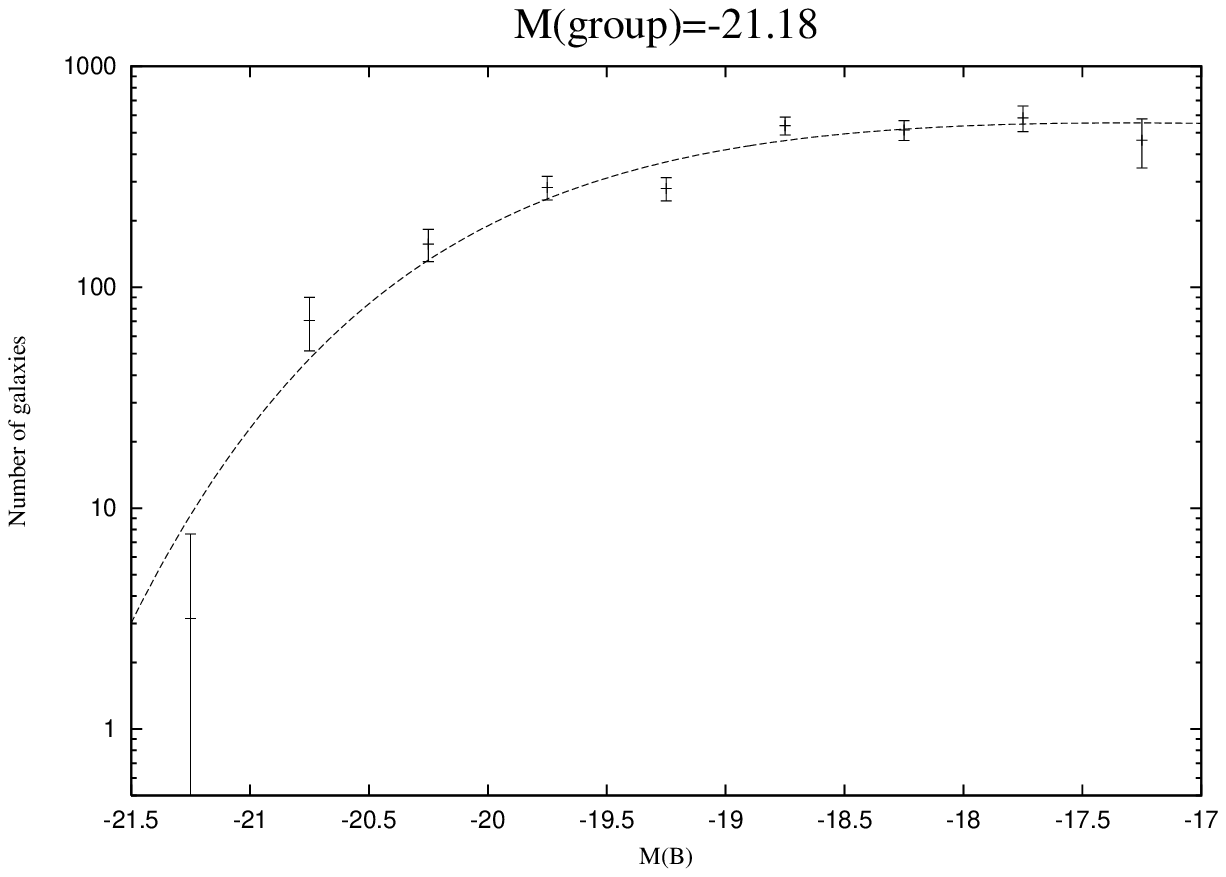}{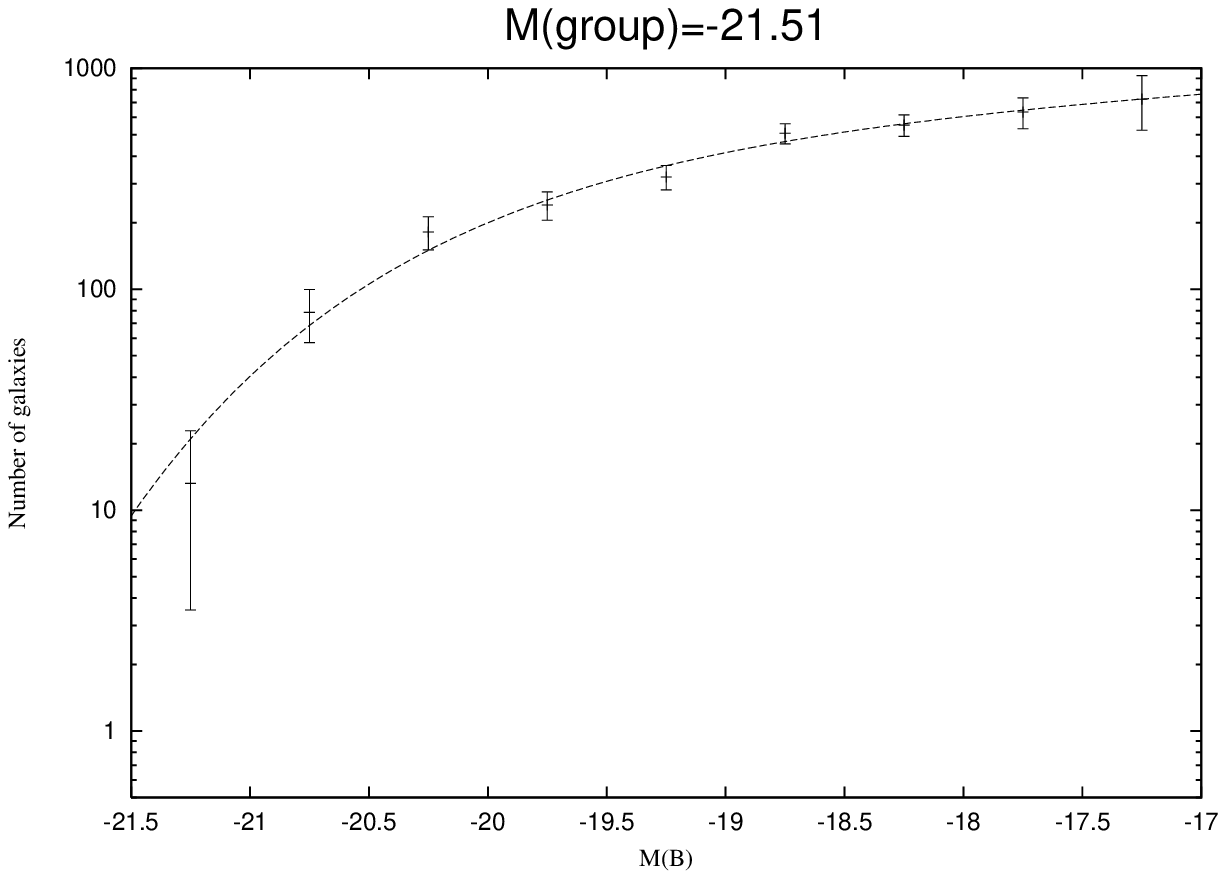} \\
\plottwo{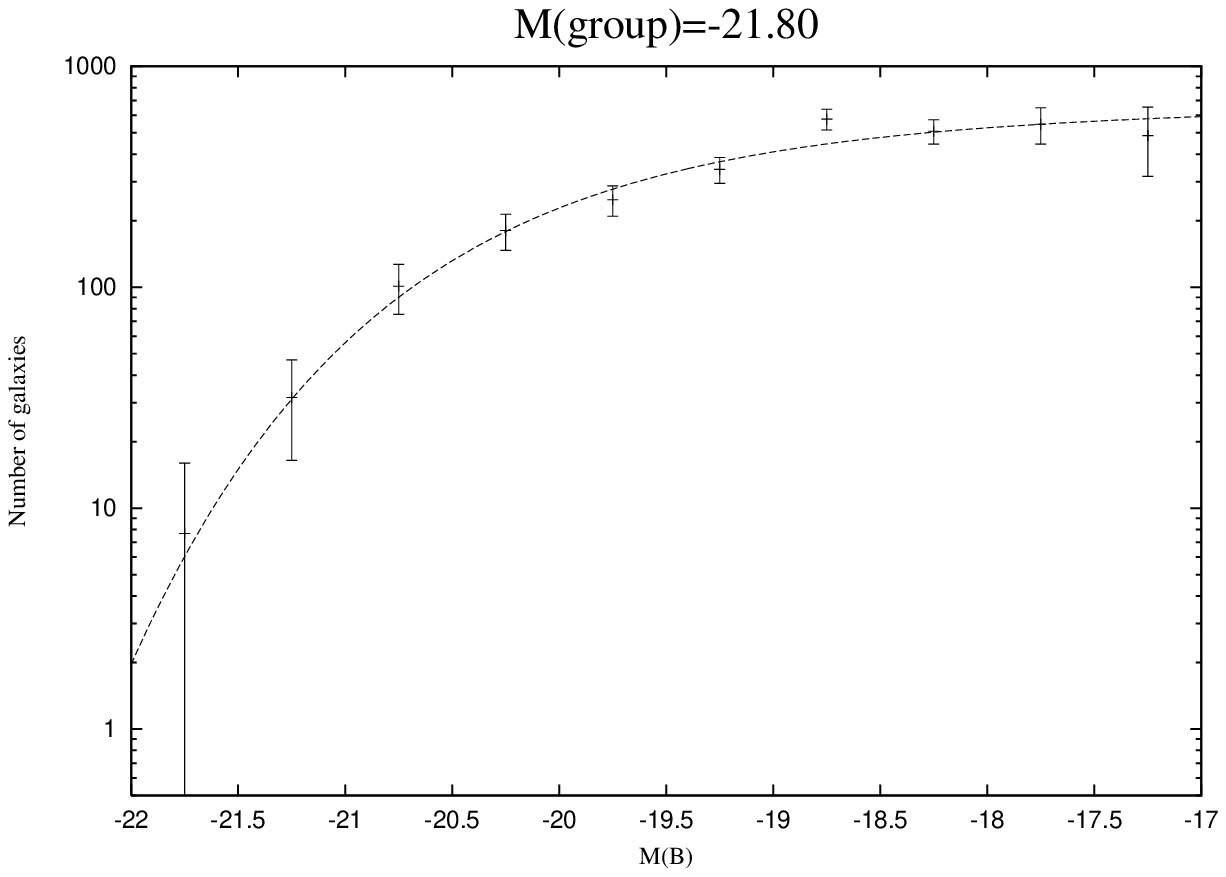}{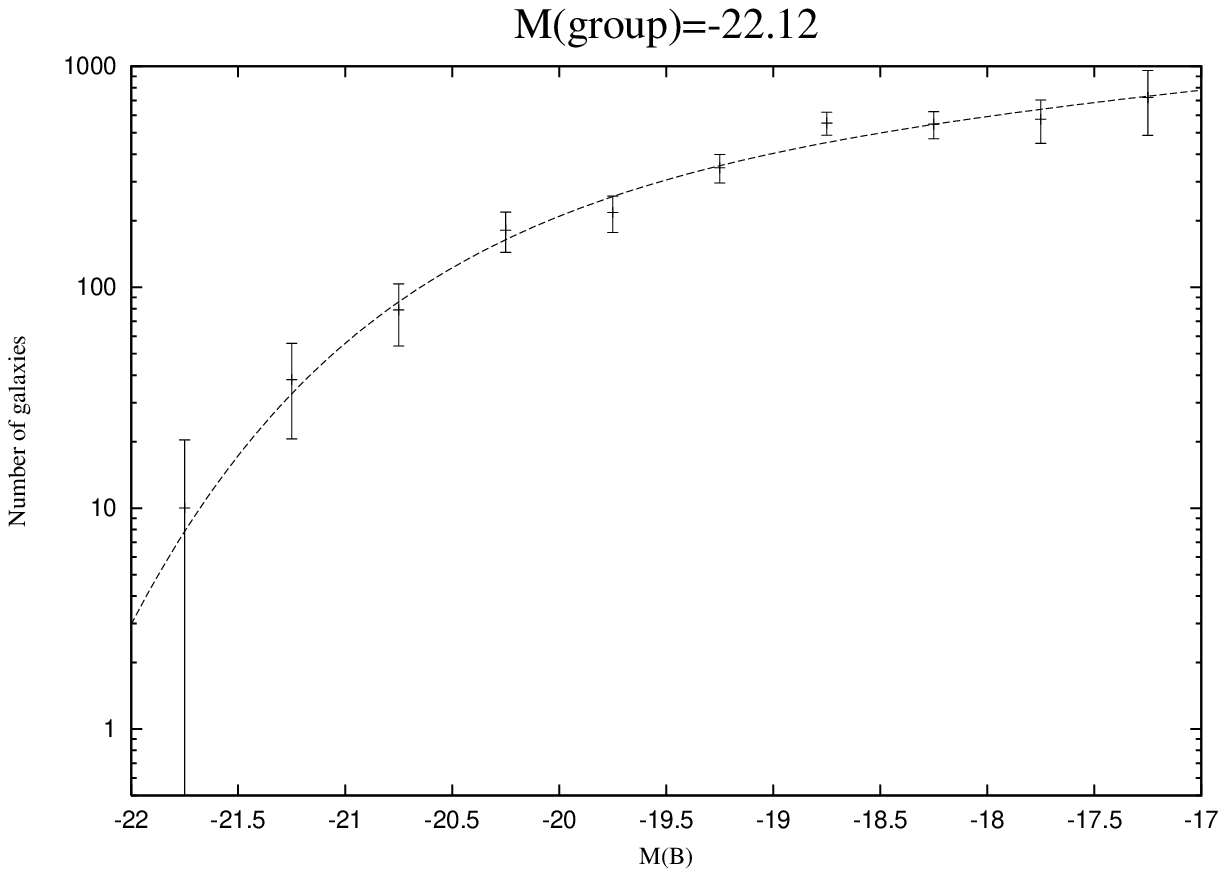}
\caption{Composite luminosity functions for galaxies in the six lowest luminosity
bins of Table 1 for the all galaxies sample.}
\end{figure}

\clearpage

\begin{figure}
\epsscale{0.95}
\plottwo{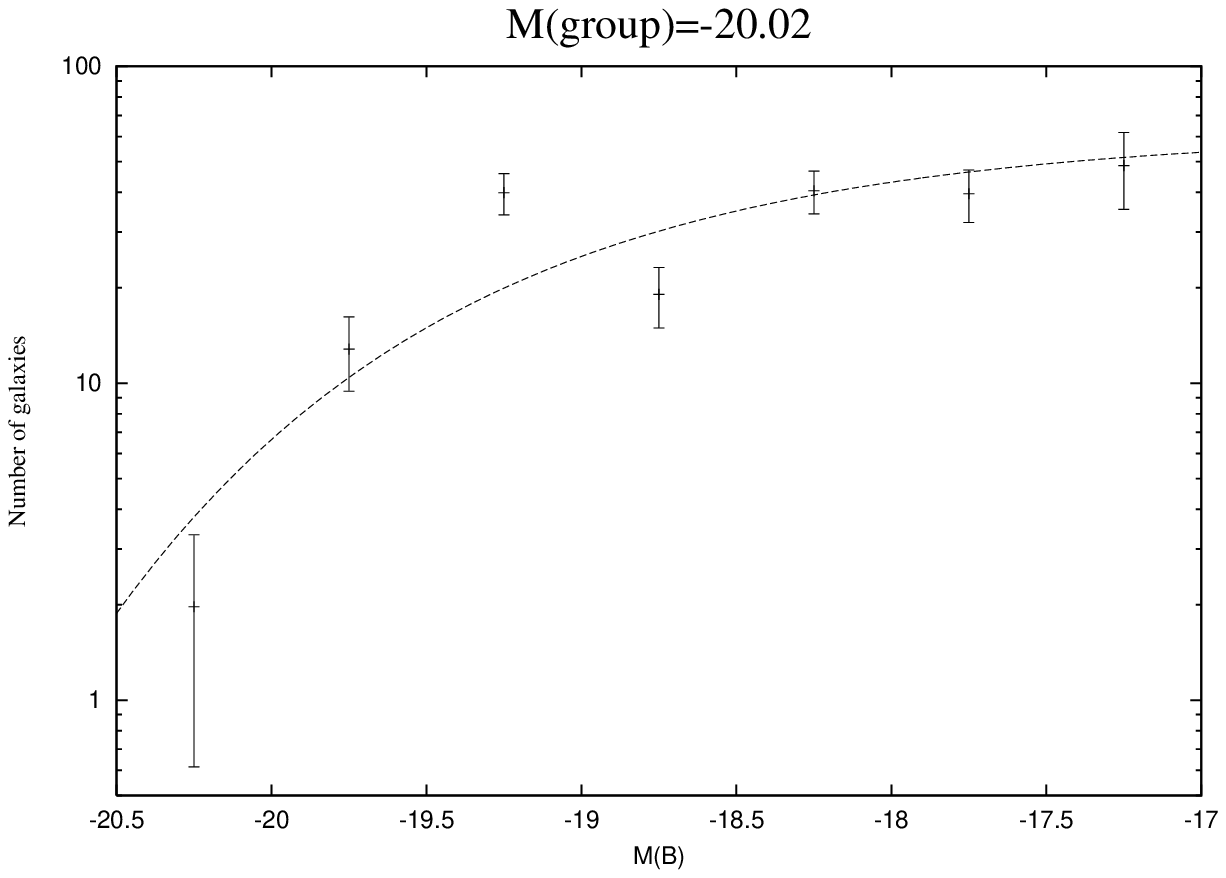}{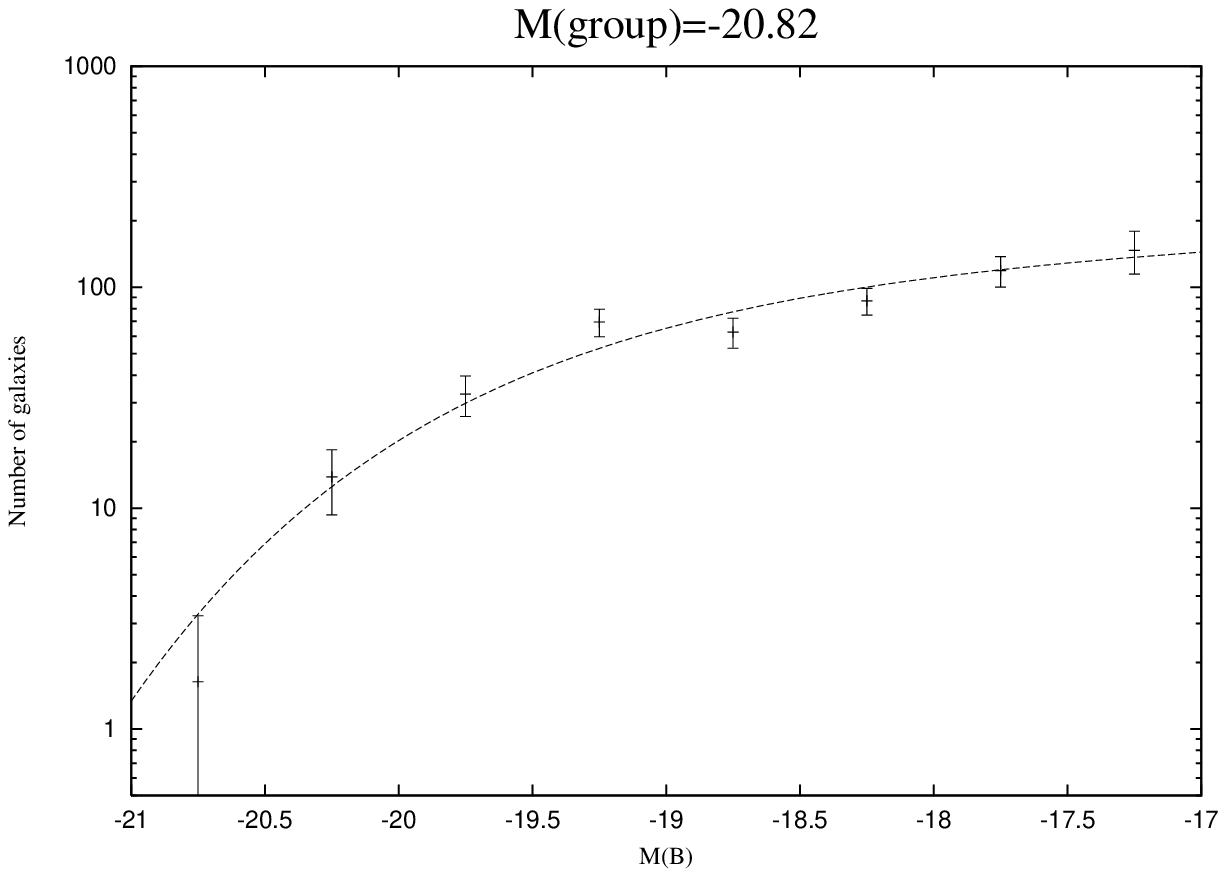} \\
\plottwo{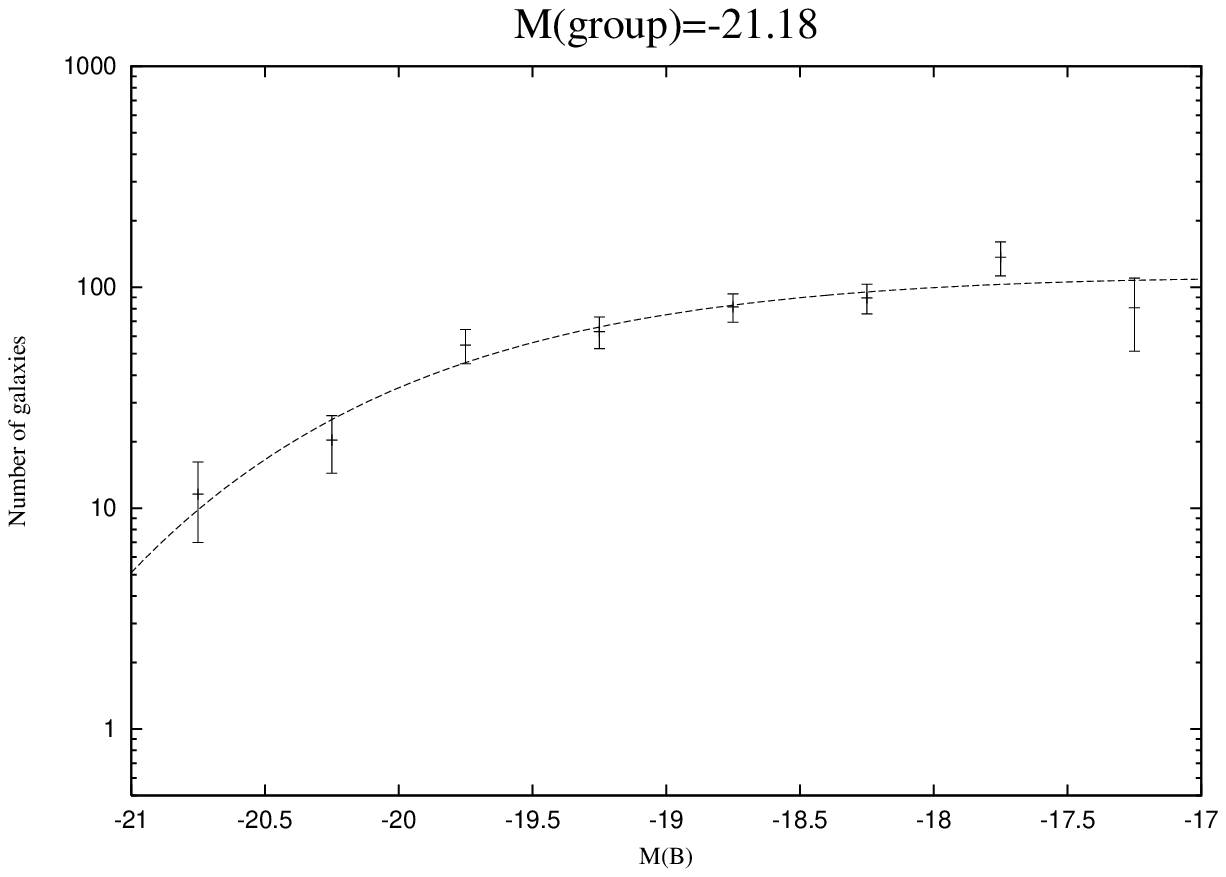}{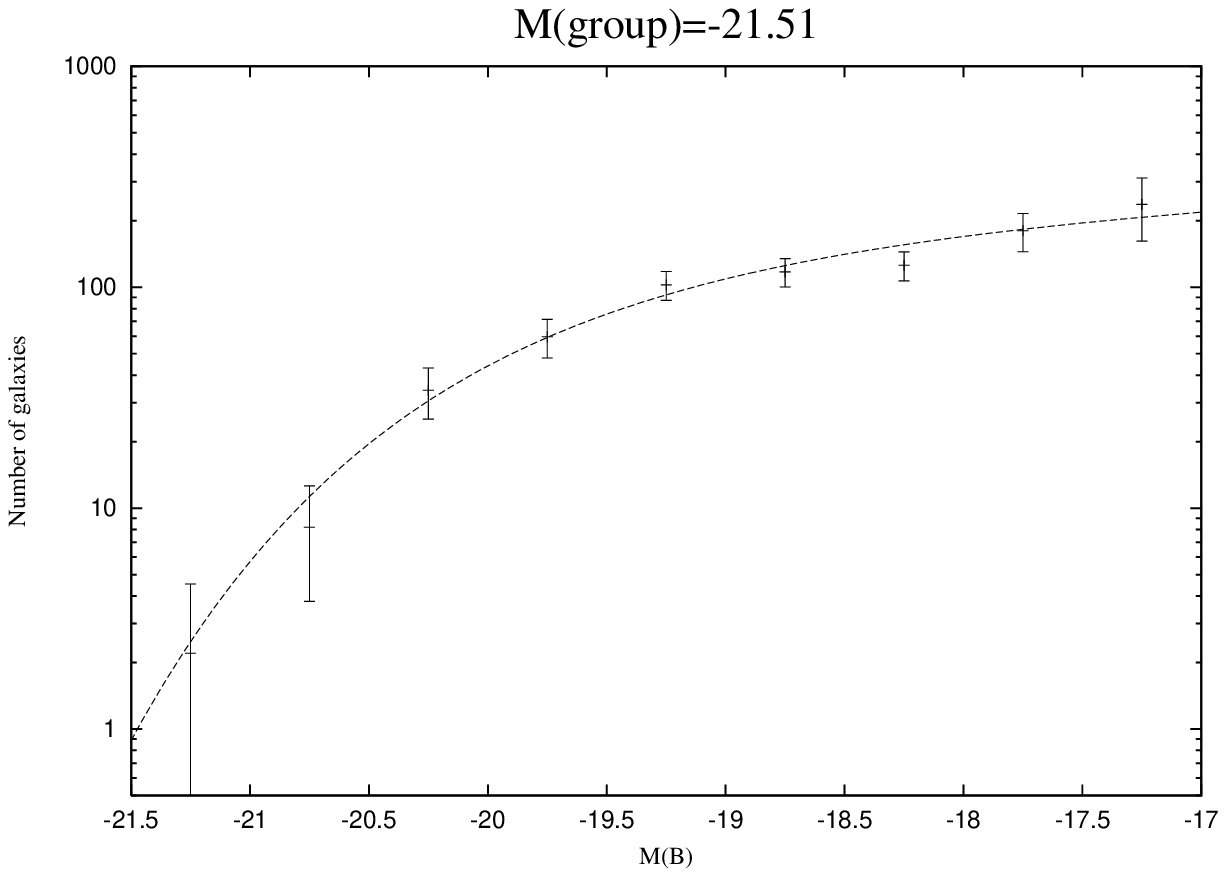} \\
\plottwo{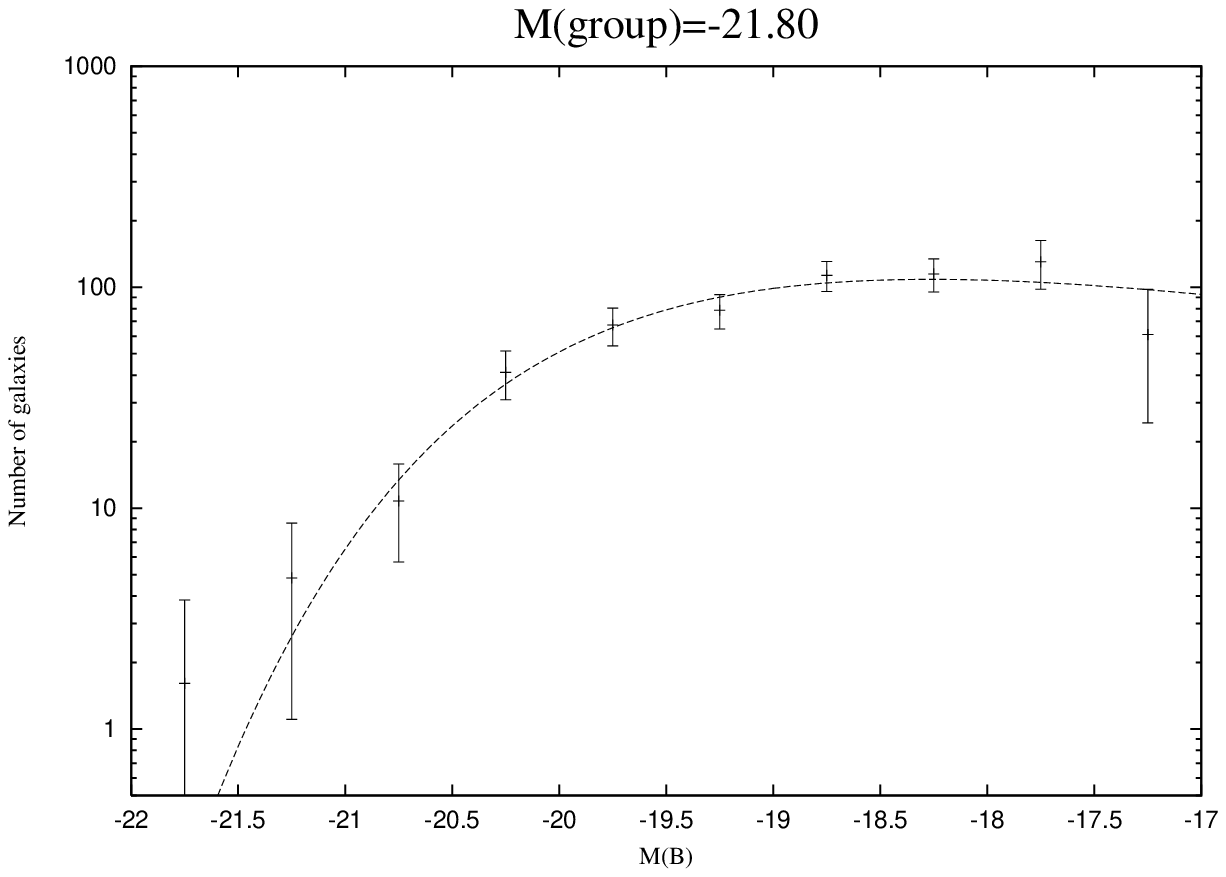}{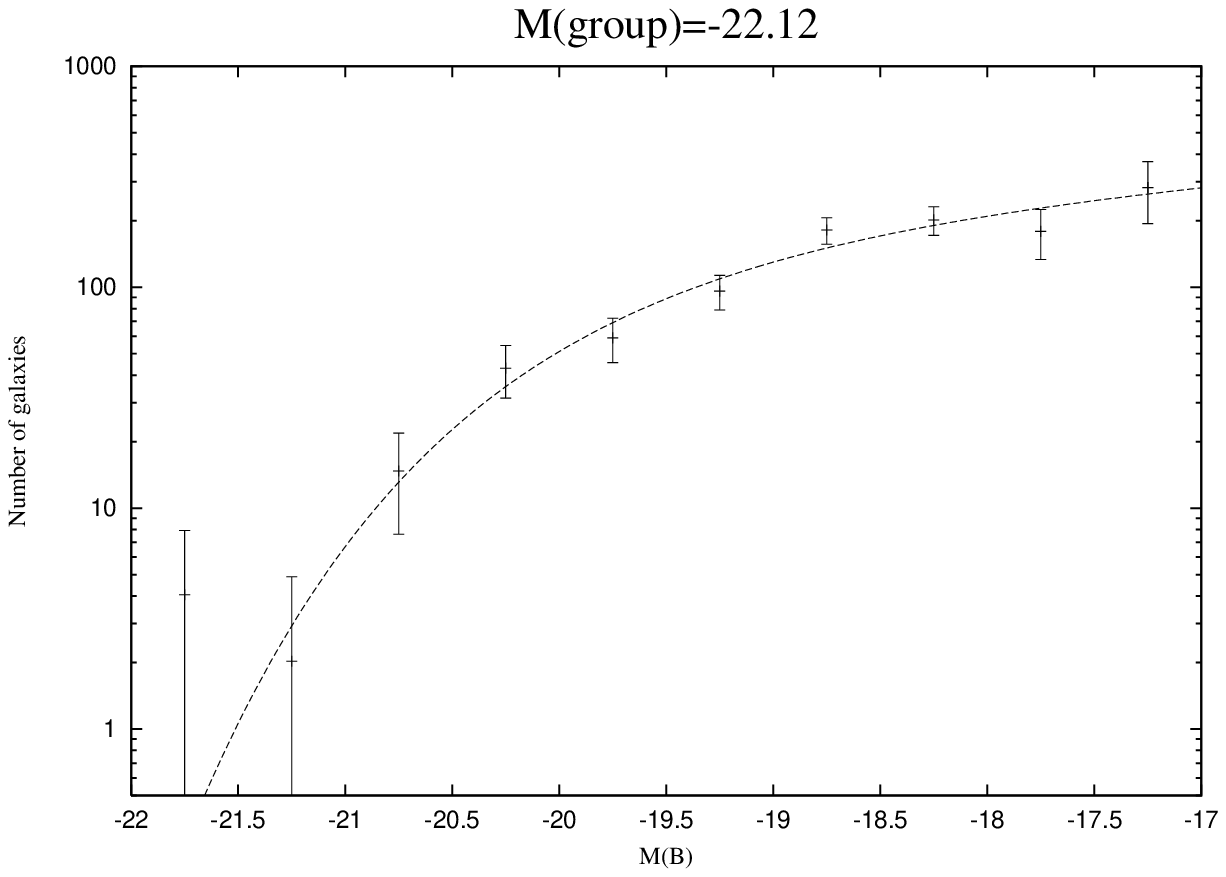}
\caption{Luminosity functions for galaxies in the blue sample.}
\end{figure}

\clearpage

\begin{figure}
\epsscale{0.95}
\plottwo{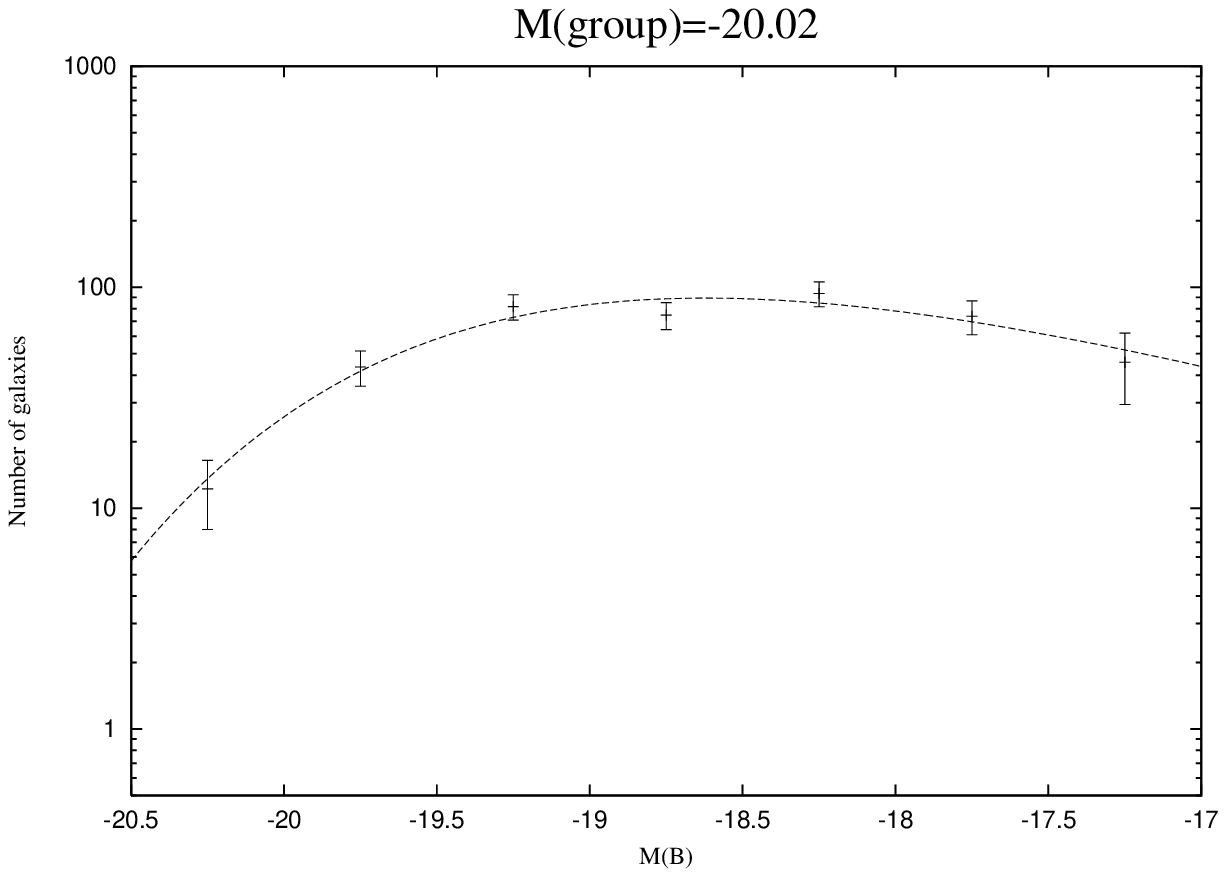}{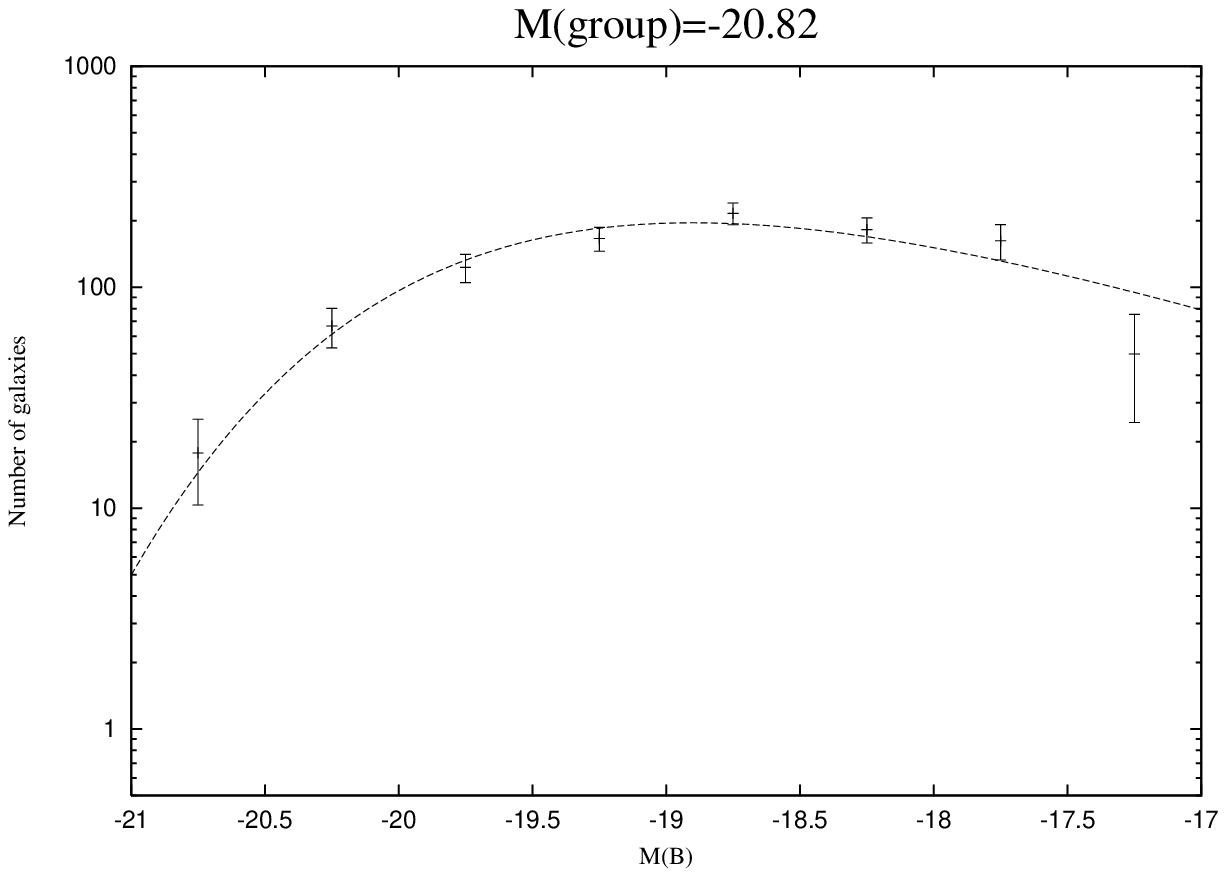} \\
\plottwo{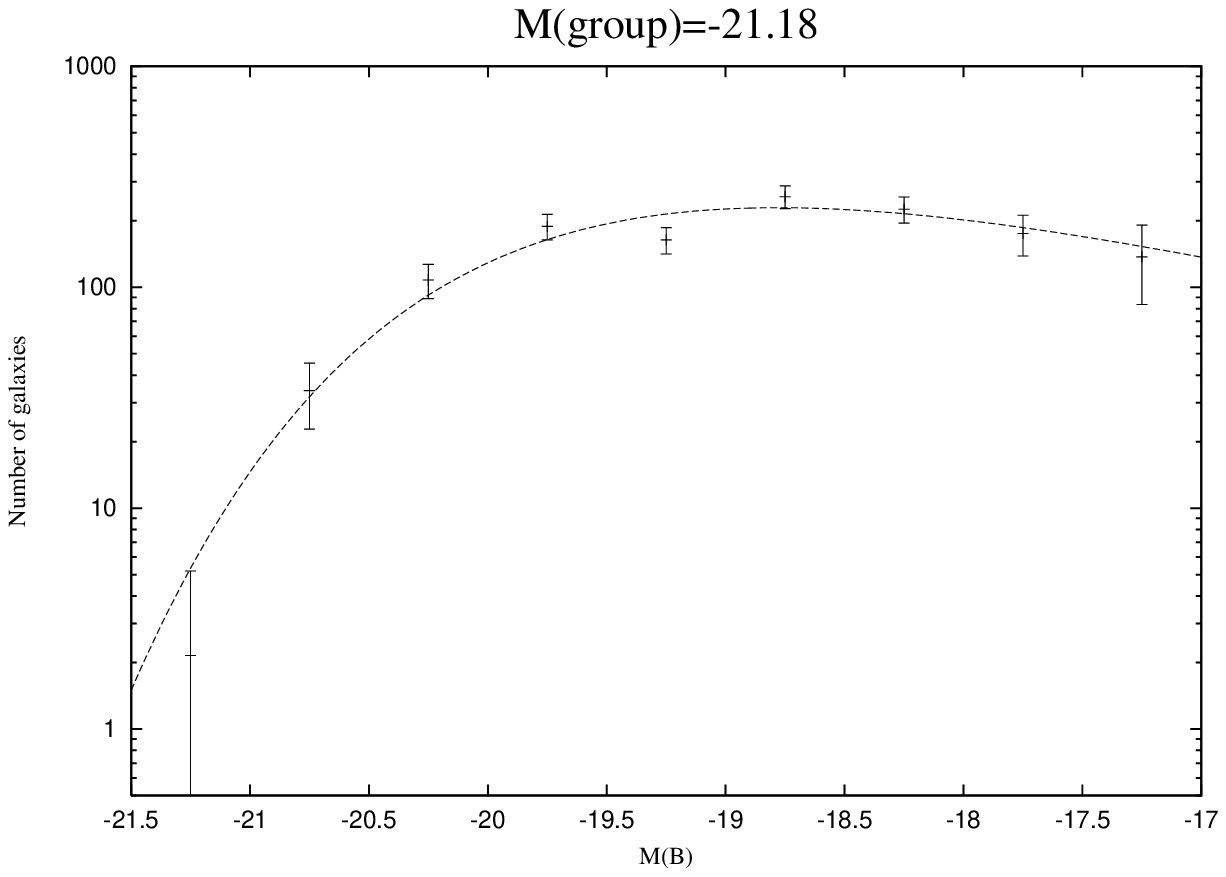}{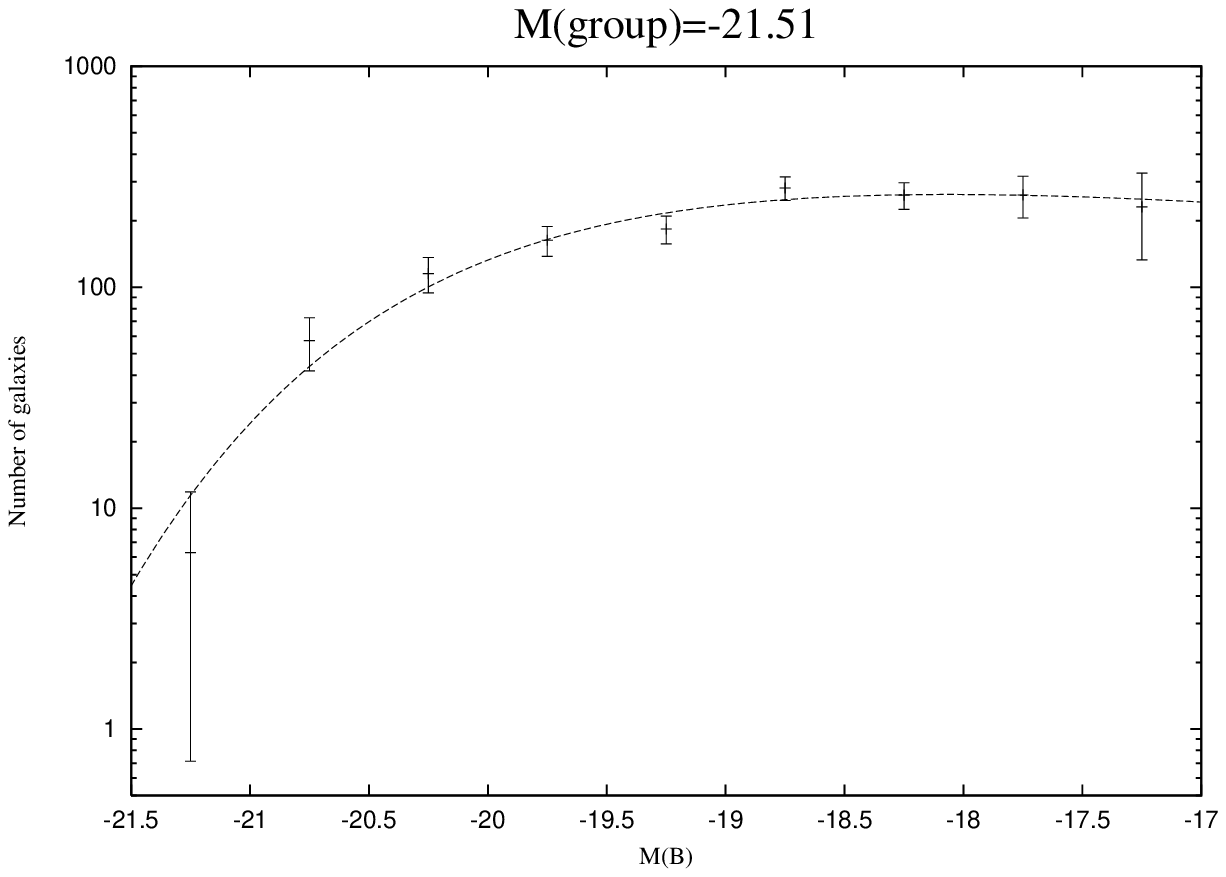} \\
\plottwo{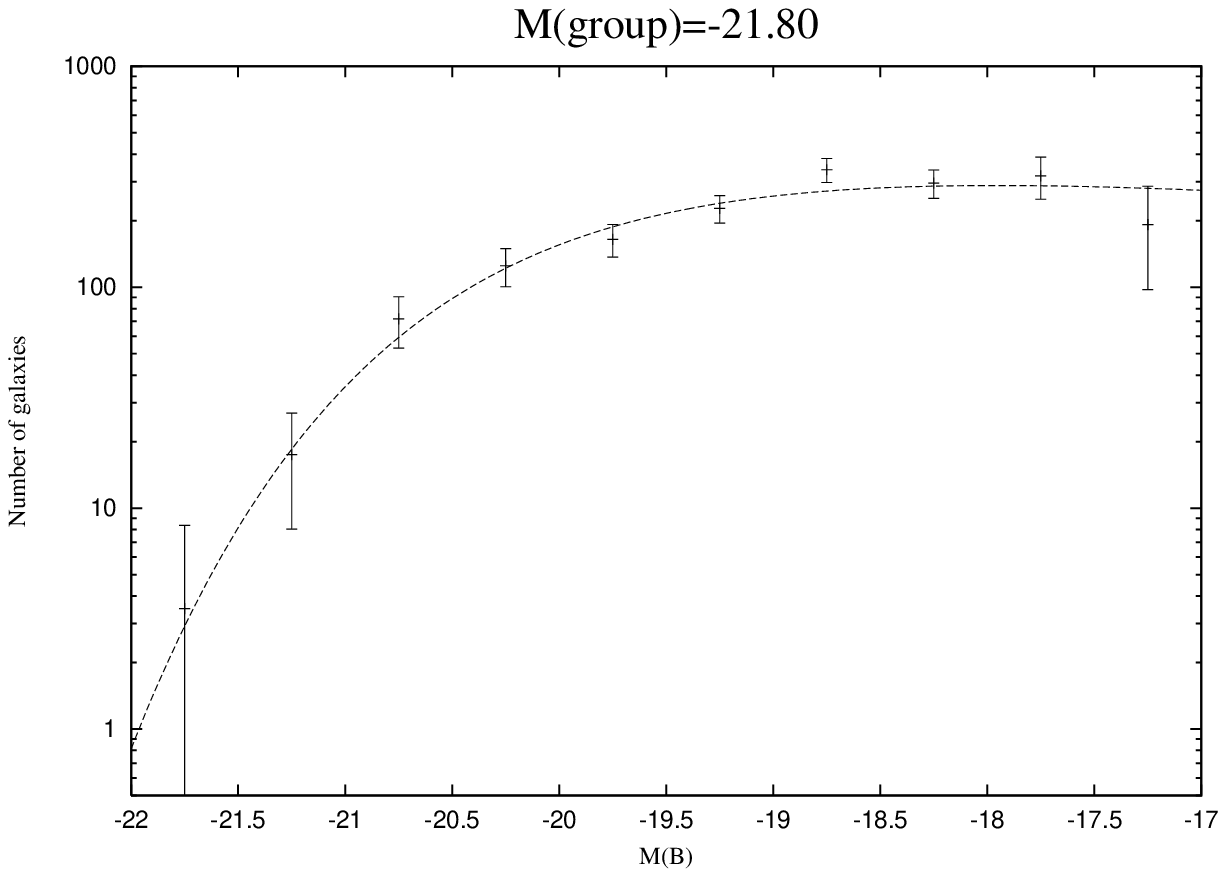}{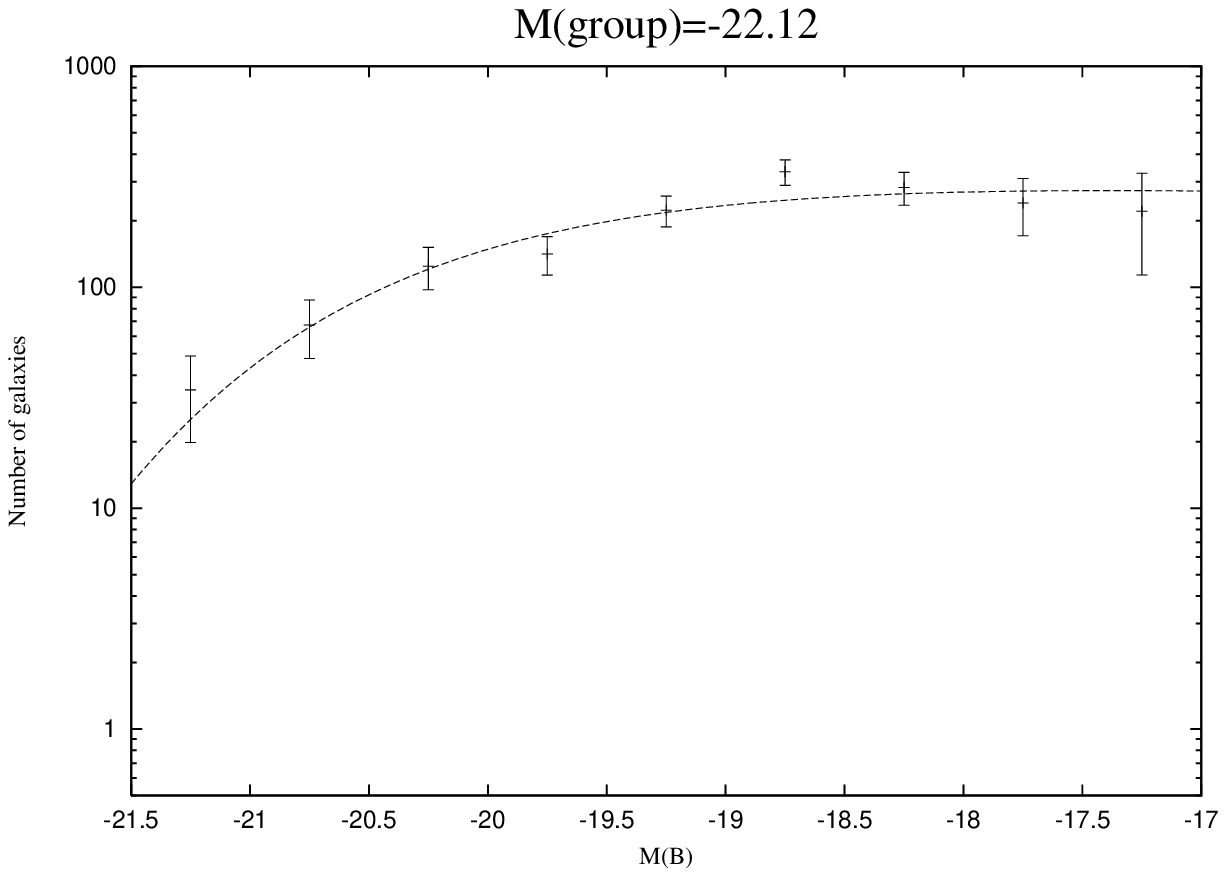}
\caption{Luminosity functions for galaxies in the red sample.}
\end{figure}

\clearpage

\begin{figure}
\epsscale{0.80}
\plotone{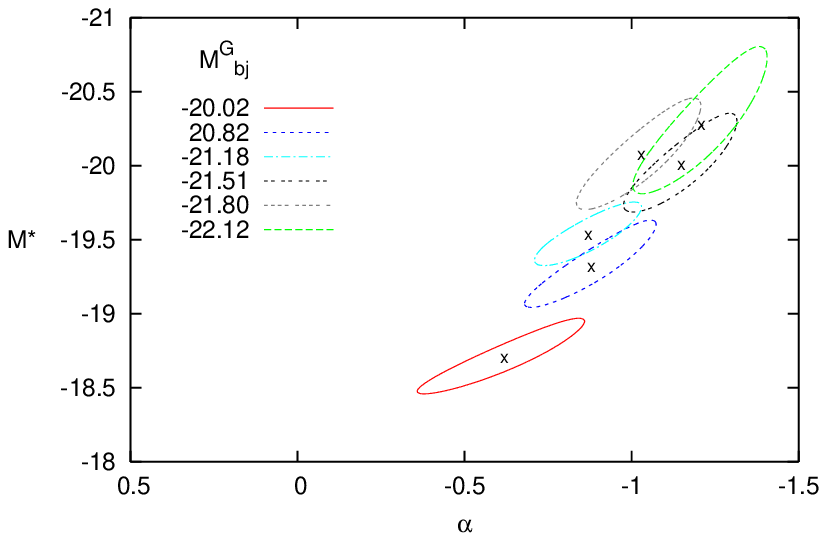}\\
\plotone{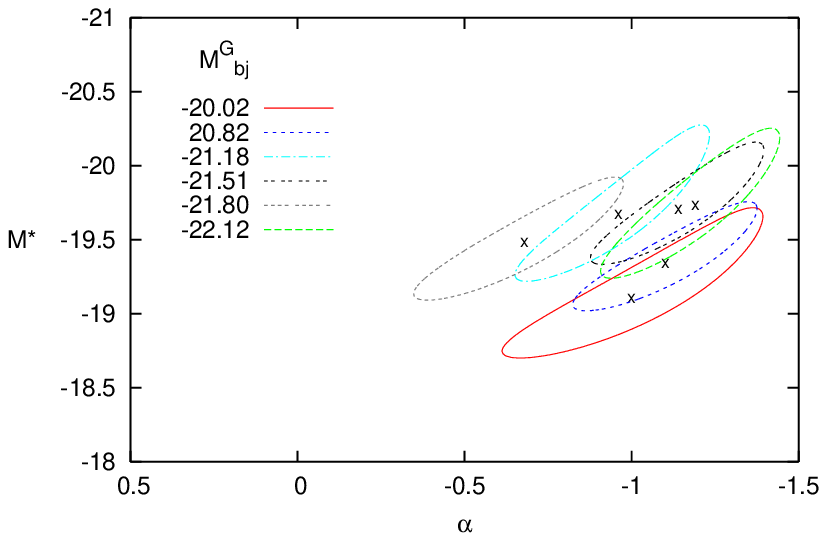}\\
\plotone{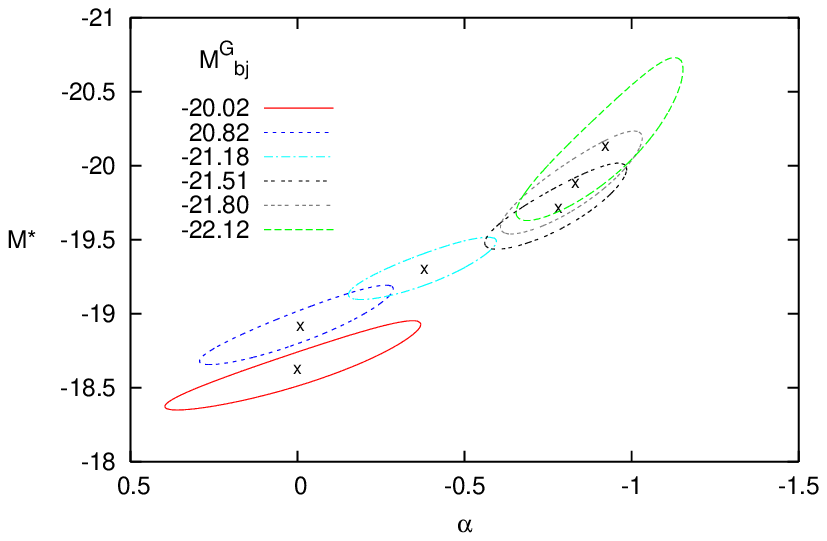}
\caption{Error ellipses in $M^*$  and $\alpha$ (for all galaxies) for the six fainter
group luminosity bins as shown in the key: the upper panel shows the ellipses for the
all galaxies sample, the middle panel the blue galaxies sample and the bottom panel
ellipses for the red galaxies sample}
\end{figure}
\clearpage

\begin{figure}
\plotone{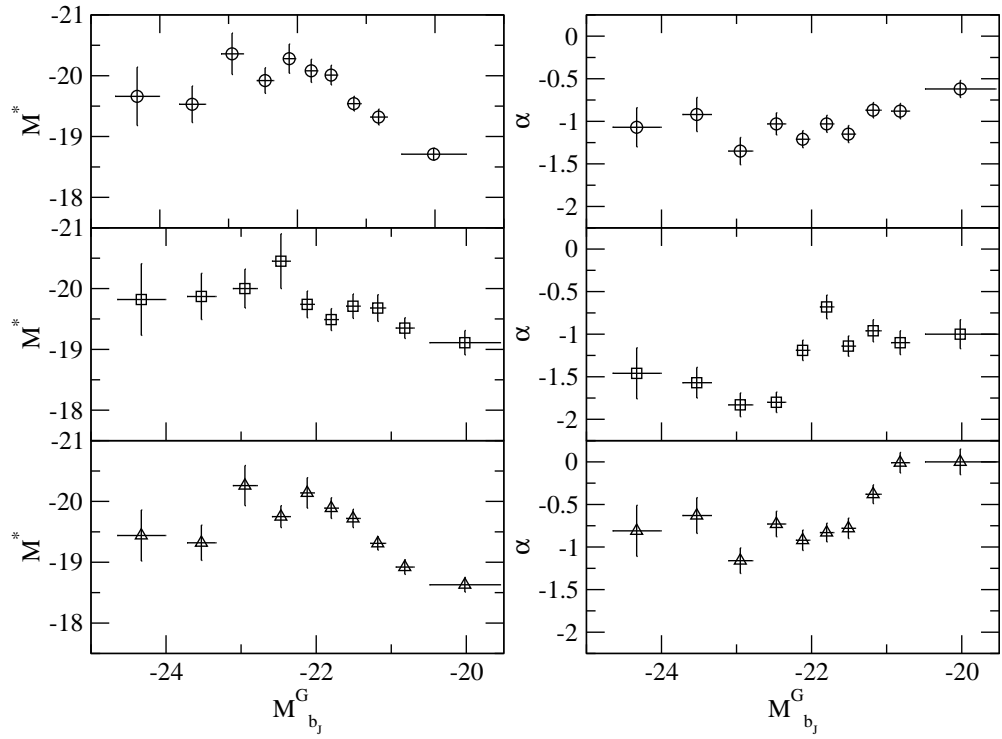}
\caption{Observed trends in $M^*$  and $\alpha$ (squares) as a function of group luminosity
for all galaxies (circles), blue galaxies (squares) and red galaxies (triangles)}
\end{figure}


\end{document}